\documentstyle[12pt]{article}
\renewcommand{\d}{\delta}
\renewcommand{\b}{\beta}
\newcommand{\g}{\gamma}

\newcommand{\s}{\sigma}
\newcommand{\la}{\lambda}
\renewcommand{\a}{\alpha}
\begin{document}
\title{Easy Control over Fermionic Computations}
\author{Yuri Ozhigov
\\[7mm]
{\it Institute of Physics and Technology,}\\
{\it Russian Academy of Sciences,}\\ 
{\it Nakhimovsky pr. 34,}\\ 
{\it Moscow, 117218, Russia,}\\
{\it e-mail: ozhigov@ftian.oivta.ru }}
\date{}
\maketitle
\begin{abstract} Quantum fermionic computations on occupation numbers proposed in quant-ph/0003137 are studied. 
It is shown that a control over external field and tunneling would suffice to
fulfill all quantum computations without valuable slowdown in the framework of
such model when an interaction of diagonal type is fixed and permanent. 
Substantiation is given 
through a reduction of some subset of this model to the conventional language of quantum 
computing and application of the construction from quant-ph/0202030.
\end{abstract}
\section{Introduction and background}

Quantum computing is an unprecedented examination of the modern physics because 
it requires a level of control over microscopic sized objects which has never 
been reached artificially. While a mathematical theory of quantum computations 
is well developed its physical implementation represents a serious challenge to 
our understanding of Nature. This is why it is important to look for its 
simplest possible realization which would depend only on fundamental principles 
of quantum mechanics and contain minimal technological difficulties. Two 
requirements can be formulated for such scheme: adequate description of states 
forming a basis of computational Hilbert space, and realistic method of control. 
Conventionally a computational element - qubit is represented as some 
characteristic like spin, charge or position of some elementary particle. This 
approach works well for one isolated qubit. For a system of many qubits it meets 
a serious difficulty. It comes from a fundamental physical principle of identity 
of elementary particles of the same type\footnote{I express my thanks to Sergei Molotkov 
who attracted my attention to this problem and referred me to the paper \cite{KB}. I am 
also grateful to Alexander Tsukanov and Alexander Kokin for useful comments to the 
preliminary version of this paper.}. 
To control a computation we must be able to address a separated qubit whereas different 
particles are undistinguishable according to the principle of identity. Of course, we can 
distinguish particles by their spatial positions placing them fairly far one from another 
but in this case it will be difficult to keep them in entangled state that is necessary 
for quantum computations. A solution of this dilemma was found by Kitaev and Bravyi, who proposed to 
use a Fock space of occupation numbers to the description of quantum computations 
(see \cite{KB}). It uses in fact a natural identification of qubits with energy levels 
in Fock space where one is treated as occupied level and zero as free level. This 
approach gives 
a universal quantum computing for a high price: it requires a control over 
external field, tunneling, diagonal interaction and changes of numbers of 
particles (contact with a superconductor) that is to control over coefficients 
$\a ,\b , \g$ in (\ref{ham}) and over additional summand $\d a_k^+ a_j^+ +\d^* a_k a_j. $

 We shall see how to reduce this price 
using an idea of fixed and permanent interaction. To do this we need two things: 
assumption 
that the corresponding Hamiltonian consists only of 
diagonal and tunneling summands and modified correspondence between states in 
occupation number representation and Hilbert spaces. Then to fulfill any quantum 
computation it is required to control over only 
an external field and tunneling. Such king of control is in principle realizable 
by lasers. The main scheme is presented in the section 3 and is in fact based on 
the idea of computational model with fixed permanent interaction proposed in works 
\cite{Oz, OF} adapted to the language of Fock spaces. The next section contains a short 
description of the 
occupation numbers formalism. 

\section{Formalism of occupation numbers}

In this paper we shall consider a system of $n$ identical fermions. At first 
make nonphysical assumption that they can be reliably distinguished. Then its 
state belongs to the Hilbert space of all states with the basis $\psi (r_1 , r_2 
,\ldots , r_n )$ $=\psi_{j_1} (r_1 )\psi_{j_2} (r_2 )\ldots\psi_{j_n} (r_n )$ where 
$\{\psi_j \}$ are some basis for one particle states, where $j_s$ belongs to the general 
set of indices $1,2,\ldots ,J$, $r_j$ includes spatial and spin 
coordinates. A choice of basis means that the system after measurement can be 
found in one of the basic states. In a real system of identical particles they 
cannot be distinguished. Hence any basic state must contain all summands of the 
form $\psi_{j_1} (r_1 )\psi_{j_2} (r_2) )\ldots\psi_{j_n} (r_n )$ with some 
factors. Such state must change the sign after permutation of every two 
fermions and it is convenient to assume that a basic state for $n$ fermions 
system is given by 
\begin{equation}
\Psi=\frac{1}{\sqrt{N!}}\left|
\begin{array}{ccccc}
&\psi_{j_1} (r_1 ) &\psi_{j_1} (r_2)&\ldots &\psi_{j_1} (r_n )\\
&\vdots &\vdots &\vdots &\vdots\\
&\psi_{j_n }(r_1 ) &\psi_{j_n} (r_2 ) &\ldots &\psi_{j_n} (r_n )\\
\end{array}
\right| .
\label{state}
\end{equation}

This state may be considered as a situation when only states 
$\psi_{j_s}$ for $s=1,2,\ldots ,n$ are occupied by particles from our system and all 
others $\psi_k$ for $k\in\{ 1,2,\ldots , J\}$ which have not form $j_s$ are free. If 
$\psi$ with indices denotes eigenvectors of one particle Hamiltonian we speak about 
occupied or free energy levels, but generally speaking $\psi_k$ may form arbitrary basis 
in the space of states for one particle. A state of the form (\ref{state}) may be 
represented as a symbol $|\bar n_\Psi\rangle =|n_1 ,n_2 ,\ldots ,n_J \rangle$ where 
$n_k$ is one if $k$th energy level is occupied and zero if it is free. This is a 
representation of states of fermionic ensemble in form of occupation numbers. Such 
vectors $\bar n$ form a basis of Fock space and the general form of state of our system 
will be $\sum\limits_{\bar n}\la_{\bar n}|\bar n\rangle$ with amplitudes $\la$. 

An operator of annihilation $a_j$ of a particle on $j$th level and its conjugated $a_j^+$ 
(creation) are defined by 
$a_j |n_1 ,\ldots ,n_J \rangle =\d_{1,n_j}(-1)^{\s_j}| n_1 ,\ldots
 ,n_{j-1},n_j -1,n_{j+1},\ldots ,n_J \rangle$ 
where $\s_j =n_1 +\ldots +n_j$. They possess the known commutative relations: 
$a_j^+ a_k +a_k a_j^+ =\d_{j,k}$, $a_j a_k +a_k a_j=a_j^+a_k^++a_k^+a_j^+=0$.

Assume that any interaction in Nature goes between no more than two particles. 
Hence any interaction in many-particle system may be expanded into the sum of 
one and two particles interactions of the form $H=H_{one}+H_{two}$ with the 
corresponding potentials $V_1 (r)$ and $V_2 (r,r')$. Each of them can be 
represented by operators of creations and annihilations as 
$H_{one}=\sum\limits_{k,l}H_{k,l} a^+_k a_l$, $\ \ H_{two}=
\sum\limits_{k,l,m,n}H_{k,l,m,n}a^+_l a^+_k a_m a_n$ where 
$$
\begin{array}{ll}
&H_{k,l}=\langle\psi_k |\ H_{one}\ |\psi_l \rangle =\int \psi^*_k (r) V_1 (r)\psi_l (r)dr,
\\
&H_{k,l,m,n}=\langle\psi_l ,\psi_k\ |H_{two}\ |\ \psi_m \psi_n \rangle
 =\int\psi^*_k (r)\psi^*_l (r') V_2 (r,r')\psi_m (r)\psi_n (r') drdr'.
\end{array}
$$
Hence given potentials of all interactions and all basic states $\psi_i$ we can in 
principle obtain its representation in terms of creations and annihilations, e.g. in 
the language of occupation numbers. 

Consider an ensemble with Hamiltonian of the form $H=\sum_i H^i_{ext. f.}+\sum_{i,j}
(H^{i,j}_{diag.}+H^{i,j}_{tun.})$ where Hamiltonians of external field, diagonal 
interaction and tunneling are represented by means of operators of creations and 
annihilations by 
\begin{equation}
\begin{array}{lll}
&H^i_{ext.f.} &= \a_i a^+_i a_i ,\ \ \ \ \a_i\in{\rm R},\\
&H^{i,j}_{diag.} &= \b_{i,j} a^+_i a_i a^+_j a_j ,\ \ \ \ \b_{i,j}\in{\rm R},\\
&H^{i,j}_{tun.} &= \g_{i,j} a^+_i a_j +\g_{i,j}^* a^+_j a_i .
\end{array}
\label{ham}
\end{equation}

Note that it would not be easy to implement a control over diagonal part of Hamiltonian. 
Assume that the diagonal interaction is fixed and acts permanently whereas external field 
and tunneling are subjects of control. Then it is possible to fulfill every quantum 
computation. This type of control seems to be fairly realistic because a tunneling may 
be controlled by laser impulses. 

\section{Computation controlled by tunneling}

Instead of simple correspondence between occupation numbers and Hilbert spaces described 
above 
we now establish another correspondence that makes possible to transfer 
a universal computing with fixed permanent interaction (\cite{Oz, OF}) to the language of 
fermionic computing in Fock space of occupation numbers.

Let us fix some partitioning of all energy levels to two equal parts and choose 
some one-to-one correspondence between them. Say we can consider $k$th level 
down from Fermi bound $\epsilon_F$ and agree that it corresponds to $k$th level 
up from $\epsilon_F$. We shall denote $j$th level down from Fermi bound by 
ordinary letter and $j$th level up from Fermi bound by $j'$. Call the first 
level $j$th lower level and the second one $j$th upper level. Fock space $\cal 
F$ can be represented as ${\cal F}={\cal F}_1 \bigotimes {\cal F}_2 
\bigotimes\ldots\bigotimes {\cal F}_k$ where each ${\cal F}_j$ corresponds to 
$j$th pair of the corresponding energy levels. Consider a subspace $F_j$ in 
${\cal F}_ j$ which is spanned by two following states. The first one is: 
"$j'$th level is occupied and $j$th level is free", the second is "$j$th level 
is occupied and $j'$th is free". Denote them by $|1\rangle_j$ and 
$|0\rangle_j$ correspondingly. We shall deal with subspace $F=F_1 \bigotimes F_2 
\bigotimes\ldots\bigotimes F_k$ in Fock space ${\cal F}$. Now determine a 
function $\theta$ that maps our Hilbert space $\cal H$ to  $F$ by the following 
definition on basic states: $\theta (|\xi_1 ,\xi_2 \ldots\xi_n \rangle )=|\xi_1 
\rangle_1 \bigotimes |\xi_2 \rangle_2 \bigotimes\ldots\bigotimes |\xi_n \rangle_n$ where 
all 
$\xi_j$ are ones and zeroes. Thus $\theta$ establishes 
unconventional correspondence between Hilbert and Fock spaces (see Figure 1).

One qubit state in Hilbert space corresponds to two qubits state in conventional 
assignment of qubits for Fock space - each occupation number to each qubit. 
But we shall see that this assignment answers to the task of control over 
computation better than conventional assignment. 

Now the door is open for representation of unitary transformations in Hilbert 
space required for quantum computing by transformations in Fock space. Consider Hermitian 
operator $H$ in one-dimensional Hilbert space $\cal H$. It has the form $H_0
+H_1$ where
$$
H_0 =\left(
\begin{array}{ccc}
&d_1 &0\\
&0 &d_2
\end{array}
\right) ,
H_1=\left(
\begin{array}{ccc}
&0 &d\\
&\bar d &0
\end{array}
\right) .
$$

\setlength{\unitlength}{0.6pt}
\begin{picture}(720,420)(0,90)
\put(100,500){$\xi\in{\cal H}$}
\put(247,500){$|0\rangle_2$}
\put(332,500){$|1\rangle_2$}
\put(520,500){$|010\rangle$}
\multiput(200,230)(0,36){7}{\multiput(0,0)(17,0){10}{\line(1,0){7}}\multiput(260,0)(15,0){14}{\line(1,0){5}}}
\put(255,410){\circle{10}}
\put(340,266){\circle{10}}
\put(500,374){\circle{10}}
\put(536,266){\circle{10}}
\put(572,446){\circle{10}}
\put(340,410){\circle*{10}}
\put(500,302){\circle*{10}}
\put(536,410){\circle*{10}}
\put(255,266){\circle*{10}}
\put(572,230){\circle*{10}}
\put(100,338){$\theta (\xi )$}
\put(200,338){\line(1,0){170}}
\put(460,338){\line(1,0){215}}
\put(700,338){$\epsilon_F$}
\put(150,230){$3$}
\put(150,266){$2$}
\put(150,302){$1$}
\put(150,338){$0$}
\put(150,374){$1'$}
\put(150,410){$2'$}
\put(150,446){$3'$}
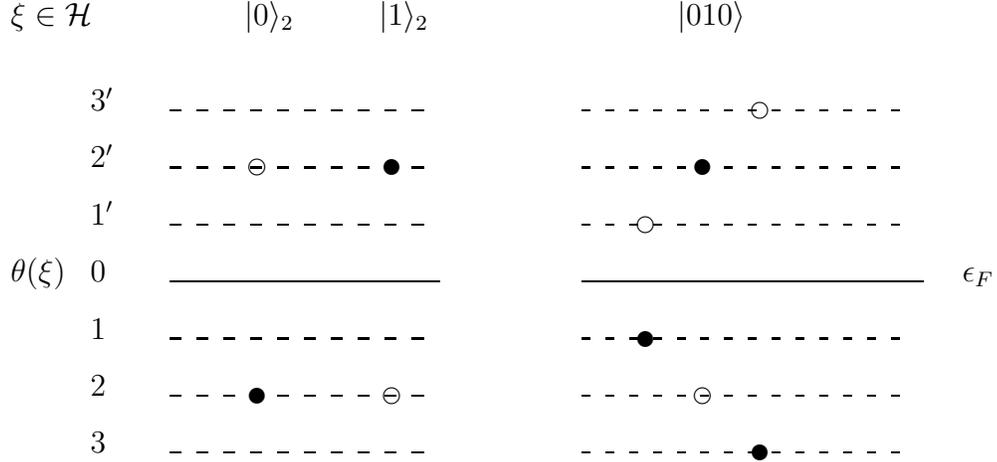
\put(150,150){Figure 1. \ \ Correspondence between Fock and Hilbert spaces}
\end{picture}

It can be straightforwardly verified that for operators $\tilde H_0 = d_1 a^+_k 
a_k +d_2 a^+_{k'} a_{k'}$ and $\tilde H_1 = d a^+_k a_{k'} +\bar d a^+_{k'} a_k$ 
(that is external field and tunneling) we have equalities $ \tilde H_i \theta 
=\theta H_i$ for $i=0,1$. Using linearity of $\theta$ we obtain $(\tilde H_0 
+\tilde H_1 )\theta =\theta H$.
Now consider one qubit unitary transformations $U$ in Hilbert space. It has the 
form $e^{-iH}$ for Hamiltonian $H$ (we choose appropriate time scale to get rid 
of Plank constant and time). Using linearity of $\theta$ and the equality 
$\theta^{-1} H^s \theta =(\theta ^{-1}H\theta )^s$ for natural $s$ we obtain 
that for every one qubit unitary transformation $U$ we can effectively find the 
corresponding Hamiltonian in Fock space containing only external field and 
tunneling that makes diagram A from the Figure 2 closed. 

Take up two qubits transformations in Hilbert space. Since all diagonal matrices 
commute, for all diagonal transformation in space ${\cal F}_k 
\bigotimes {\cal F}_j$ we can effectively find the corresponding diagonal
 transformation on the corresponding Hilbert space that makes diagram B from 
the Figure 2 closed. Note that for entangling ${\cal V}$ the transformation $V$ 
will be entangling as well. 

Now all is ready to transfer a trick from \cite{OF} with one qubit controlled 
universal computations to Fock space. Combination of diagrams from Figure 2 gives diagram from the Figure 3.

\begin{picture}(700,230)(0,150)
\put(64,330){\vector(1,0){256}}
\put(424,330){\vector(1,0){266}}
\put(64,220){\vector(1,0){256}}
\put(424,220){\vector(1,0){266}}
\put(50,234){\vector(0,1){86}}
\put(330,234){\vector(0,1){86}}
\put(410,234){\vector(0,1){86}}
\put(700,234){\vector(0,1){86}}
\put(37,275){$\theta$}
\put(317,275){$\theta$}
\put(397,275){$\theta$}
\put(687,275){$\theta$}
\put(43,213){$\cal H$}
\put(323,213){$\cal H$}
\put(403,213){$\cal H$}
\put(693,213){$\cal H$}
\put(43,323){$\tilde F$}
\put(323,323){$\tilde F$}
\put(403,323){$\tilde F$}
\put(693,323){$\tilde F$}
\put(80,340){ext. field + tunneling}
\put(480,340){$\tilde F
$ diagonal}
\put(80,200){one-qubit on $\cal H$}
\put(480,200){$\cal H$ diagonal}
\put(150,150){A}
\put(550,150){B}
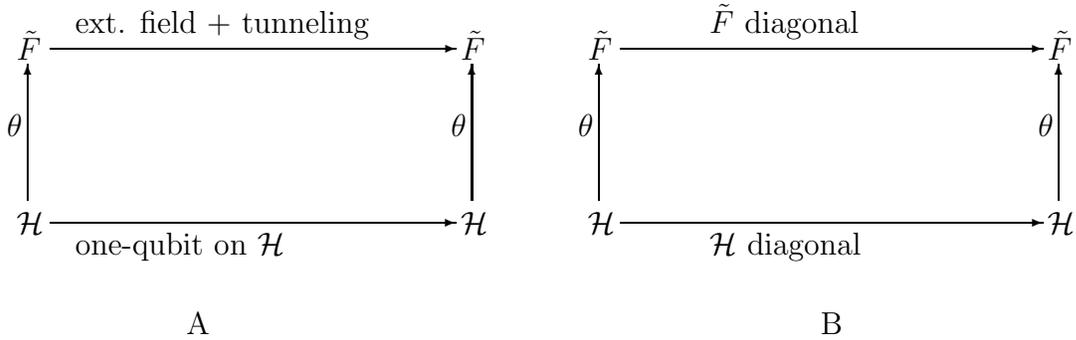
\put(120,90){Figure 2.\ \ Correspondence of transformations in }
\put(180, 65){in Fock and Hilbert subspaces. $\tilde F=F_j \bigotimes F_k .$}
\end{picture}

\begin{picture}(700,230)
\put(90,320){$\cal F$ diag}
\put(250,320){f+t}
\put(580,320){$\cal F$ diag}
\put(90,130){diag}
\put(250,130){one qubit}
\put(580,130){diag}
\put(40,300){\vector(1,0){125}}
\put(220,300){\vector(1,0){125}}
\put(550,300){\vector(1,0){125}}
\put(40,150){\vector(1,0){125}}
\put(220,150){\vector(1,0){125}}
\put(550,150){\vector(1,0){125}}
\put(30,165){\vector(0,1){115}}
\put(197,165){\vector(0,1){115}}
\put(700,165){\vector(0,1){115}}
\put(390,300){.\ .\ .}
\put(390,150){.\ .\ .}
\put(20,297){F}
\put(190,297){F}
\put(693,297){F}
\put(20,142){$\cal H$}
\put(187,142){$\cal H$}
\put(685,142){$\cal H$}
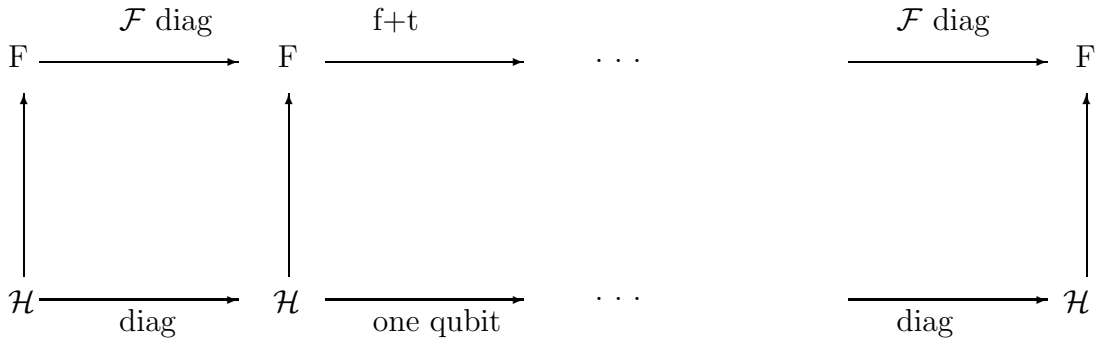
\put(100,80){Figure 3.\ \ Correspondence of computations in}
\put(180,55){Fock and Hilbert spaces}
\end{picture}

Let a diagonal part of interaction in Fock space be fixed and act permanently. 
Then we can find the corresponding diagonal interaction on Hilbert space making 
all diagonal parts of diagram from Figure 3 closed. By the result from \cite{OF} 
we can choose 1 qubit transformations implementing arbitrary quantum
 computations in Hilber space in the form represented by the lower sequence of 
the diagram. At last we can find field + tunneling control over states in Fock 
space making all diagram closed. Note that all operators of creations and annihilations 
considered in the whole 
Fock space are not local due to the factor $(-1)^{\s_j}$ depending on a given state. 
For the diagonal operators $a_j^+ a_j a_k^+ a_k$ and external fields these factors are 
compensated. A tunneling operator $a_j^+ a_{j'}$ in space $F$ brings factor $(-1)^{\s'}$ 
where $\s'=\sum\limits_{s+j}^{j'-1}n_s =j'-j$ that is independent of a given state 
$|\bar n\rangle\in F$ because for such state exactly a half of levels between $j$ and $j'$ 
are occupied. Hence a sign can be factored out of all state and ignored.

Thus we obtain a universal quantum computer on 
states in occupation numbers space controlled only by external field and tunneling.

\end{document}